\documentclass[12pt]{article}

\usepackage{bm,a4wide,here}
\usepackage{graphicx}
\usepackage{amsmath,amsxtra,amsfonts,amssymb,amstext,latexsym}

\usepackage{times}
\usepackage{mathptm}

\newcommand{\p}{\partial}
\newcommand{\wh}{\widehat}
\newcommand{\wt}{\widetilde}
\newcommand{\eps}{\epsilon}

\newcommand{\vek}[1]{{\mathbf#1}}

\newcommand{\uvek}[1]{{\wh{\mathbf#1}}}

\newcommand{\nablan}{{\nabla_{\perp}}}

\newcommand{\etal}{\textit{et al}.}

\newcommand{\ibid}{\textit{ibid}.}
\newcommand{\pdf}{{PDF}}
\newcommand{\sol}{{SOL}}
\newcommand{\lcfs}{{LCFS}}

\begin{document}

\title{\bf Intermittent transport in edge plasmas}
\author{O.~E.~Garcia, V.~Naulin, A.~H.~Nielsen, and J.~Juul Rasmussen
\\Association EURATOM-Ris{\o} National Laboratory\\
Optics and Plasma Research, OPL-128 Ris{\o}\\
 DK-4000 Roskilde, Denmark}
\date{}

\maketitle

\begin{abstract}
The properties of low-frequency convective fluctuations and
transport are investigated for the boundary region of magnetized
plasmas. We employ a two-dimensional fluid model for the evolution
of the global plasma quantities in a geometry and with parameters
relevant to the scrape-off layer of confined toroidal plasmas.
Strongly intermittent plasma transport is regulated by
self-consistently generated sheared poloidal flows and is mediated
by bursty ejection of particles and heat from the bulk plasma in
the form of blobs. Coarse grained probe signals reveal a highly
skewed and flat distribution on short time scales, but tends
towards a normal distribution at large time scales. Conditionally
averaged signals are in perfect agreement with experimental
measurements.
\end{abstract}

It is well established that the cross field transport of particles
and heat near the edge of magnetically confined plasmas is
strongly intermittent. This is observed in a variety of devices
including linear \cite{huldetal91,ct:antar} as well as toroidal
configurations \cite{endleretal95,antonietal01}. Detailed
investigations of the spatial fluctuation structure have revealed
strong indications that the intermittent nature of particle and
heat transport is caused by localized structures in the form of
plasma ``blobs'' propagating radially far into the
scrape-off-layer (\sol) of toroidal
plasmas~\cite{ct:antar,ct:boedo,ct:terry}. It was suggested that
this is caused by a dipolar vorticity field formed by the charge
separation in a density blob due to guiding-center drifts in a
curved inhomogeneous magnetic field~\cite{ct:krash}.

In this contribution we will provide a self-consistent description
of the intermittent particle and heat flux and link it to the of
the emergence and evolution of such blob-like structures. We base
our investigations on a novel model for interchange turbulence in
slab geometry for the outboard midplane of a toroidal device
\cite{garciaetal04}. The model includes the self-consistent
evolution of the full profiles in the edge/SOL. A local
Boussinesq-like model, where the ``background" profile is
separated from the fluctuations, fails to provide a realistic
description. The geometry comprises distinct production and loss
regions, corresponding to the edge and \sol\ of magnetized
plasmas. The separation of these two regions defines an effective
last closed flux surface (\lcfs), though we do not include
magnetic shear in our model. In the edge region, strong pressure
gradients maintain a state of turbulent convection. A
self-regulation mechanism involving differential rotation leads to
a repetitive expulsion of hot plasma into the \sol, resulting in a
strongly intermittent transport of density and heat.

\begin{figure}
%\begin{figure}[H]
\centering
\includegraphics[width=.60\textwidth]{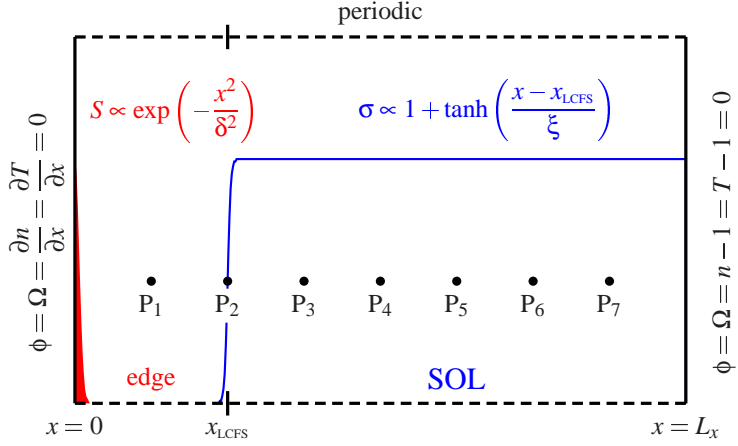}
\caption{Geometry of the simulation domain showing the forcing
region to the left, corresponding to the edge plasma, and the
parallel loss region to the right, corresponding to the scrape-off
layer. Parameters are $\delta= 8$ and $\xi=1$. Data time series
are collected at the probe positions $P_i$.} \label{fig:geometry}
\end{figure}

The model derives from the continuity equations for the electrons,
the electron temperature and the quasi-neutrality condition.
Assuming cold ions and neglecting electron inertia effects, we
obtain a three-field model for electrostatic perturbations of the
full particle density $n$, electric potential $\phi$ and electron
temperature $T$. Using slab coordinates with $\uvek{z}$ along the
magnetic field, $\uvek{x}$ in the radial and $\uvek{y}$ in the
poloidal direction we obtain~\cite{garciaetal04},
\begin{gather*}
\frac{d \Omega}{dt}  - \mathcal{C}(p) =
 \nu_{\Omega} \nabla^2 \Omega - \sigma_{\Omega} \Omega , \\
\frac{dn}{dt} + n\mathcal{C}(\phi) - \mathcal{C}(nT) =
\nu_n \nabla^2 n - \sigma_n ( n-1 ) + S_n  , \\
\frac{dT}{dt} + \frac{2T}{3}\,\mathcal{C}(\phi) -
\frac{7T}{3}\,\mathcal{C} (T) - \frac{2T^2}{3n}\,\mathcal{C}(n)
= \nu_{T} \nabla^2 T - \sigma_{T}( T-1 ) + S_{T} ,
\end{gather*}
where time is normalized by $1/\omega_{ci}$ and spatial scales by
 $\rho_s=c_s/\omega_{ci}$. The density $n$ and temperature $T$ are normalized to
 fixed characteristic
values at the outer wall. We further define the advective
derivative, the magnetic field curvature operator and the toroidal
magnetic field by
\[
\frac{d}{dt} = \frac{\p}{\p t} +
\frac{1}{B}\,\uvek{z}\times\nabla\phi\cdot\nabla , \quad
\mathcal{C} = - \zeta\,\frac{\p}{\p y}  ,  \quad
B = \frac{1}{1 + \eps + \zeta x} .
\]
The vorticity $\Omega=\nabla_{\perp}^2\phi$, the inverse aspect
ratio $\eps=a/R_0$ and $\zeta=\rho_s/R_0$ where $a$ and $R_0$ are
the minor and major radius of the device. The terms on the right
hand side of the equations describe external sources $S$, parallel
losses along open field lines through the damping rates $\sigma$
\cite{garciaetal04}, and collisional diffusion with coefficients
$\nu$. The geometry and boundary conditions are sketched in
Fig.~\ref{fig:geometry}.

In the absence of external forcing and dissipative processes the
model equations non-linearly conserves the global energy to lowest
order in $\zeta $
\[
E = \int d\vek{x}\:\left[ \frac{1}{2}\left( \nablan\phi \right)^2 +
\frac{3}{2}\,nT \right] ,
\]
where the integral extends over the whole plasma layer. We define
the kinetic energy of the fluctuating and mean components of the
flows,
\begin{equation} \label{kinenergy}
K = \int d\vek{x}\:\frac{1}{2}\left( \nabla_{\perp}\wt{\phi} \right) ,
\qquad
U = \int d\vek{x}\:\frac{1}{2}\,v_0^ 2 ,
\end{equation}
where the zero index denotes an average over the periodic direction $y$
and the spatial fluctuation about this mean is indicated by a tilde. The
linearly damped mean flows, $v_0=\p\phi_0/\p x$, does not yield any radial
convective transport and hence form a benign path for fluctuation energy.
The energy transfer rates from thermal energy to the fluctuating motions,
and from the fluctuating to the mean flows, are given respectively by
\begin{equation} \label{transfer}
F_p = \int d\vek{x}\:nT\mathcal{C}(\phi) ,
\qquad
F_{v} = \int d\vek{x}\:\wt{v}_x\wt{v}_y\frac{\p v_0}{\p x} .
\end{equation}
Note that $F_p$ is directly proportional to the domain integrated
convective thermal energy transport, while $F_v$ shows that
structures tilted such as to transport positive poloidal momentum
up the gradient of a sheared flow will sustain the flow against
collisional dissipation~\cite{ct:garcia,ct:gb,ct:naulin}.

In the following we present results from numerical simulations of
the interchange model using parameters relevant for \sol\ plasmas.
$L_x=2L_y=200$ and the \lcfs\ is located at $x_\text{\tiny
LCFS}=50$. The parameters are $\eps=0.25$, $\zeta=5 \times
10^{-4}$, and $\nu=10^{-2}$ for all fields. The parallel loss rate
of temperature is assumed to be five times larger than that on the
density and vorticity,
$\sigma_n=\sigma_{\Omega}=\sigma_T/5=3\zeta/2\pi q$, since
primarily hot electrons are lost through the end sheaths.
$\sigma_n$ and $\sigma_\Omega$ correspond to losses at the nearest
target plate over a distance of $L_{\parallel} = 2\pi R_0 q/3$
(one third of the connection length) with the acoustic speed
$c_s$, where $q=3$ is the safety factor at the edge. Finally, the
radial line-integral of the sources $S_n$ and $S_T$ equals $0.1$.
For the numerical solution the spatial resolution is  $ 512 \times
256 $ grid points in the radial and poloidal directions, and the
time span of the simulation is $2\times10^6$.

We have performed several runs with varying parameters, showing
that the qualitative behavior is robust, whereas the quantitative
results depend on the parameters and particular on the value of
the collisional diffusivities.
\begin{figure}
%\begin{figure}[H]
\centering
\includegraphics[width=.75\textwidth]{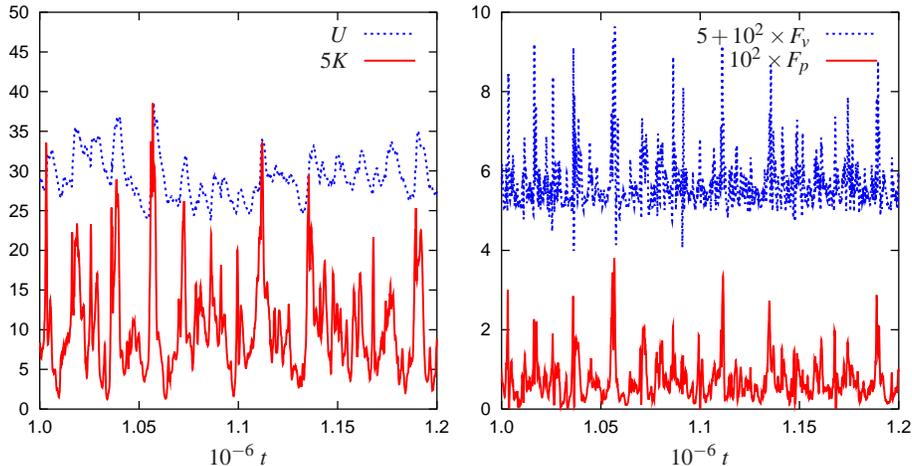}
\caption{Evolution of the kinetic energy contained by the mean $U$
and fluctuating $K$ motions [Eq.~\eqref{kinenergy}] and the
collective energy transfer terms [Eq.~\eqref{transfer}] .}
\label{fig:kinetic}
\end{figure}
The general observation is that the turbulent flux is strongly
intermittent: quite periods are interrupted by strong bursts of
particle and heat fluxes. This is correlated with the kinetics in
the fluctuations, as is shown in Fig.~\ref{fig:kinetic}. We
observe that the convective energy and thermal transport appears
as bursts during which particles and heat are lost from the edge
into the \sol\ region. As discussed in
Refs.~\cite{ct:garcia,ct:gb,ct:naulin}, this global dynamics is
caused by a self-regulation mechanism in which kinetic energy is
transfered from the fluctuating to the mean components of the
flows, and subsequently damped by collisional dissipation. The
thermal energy ejected in a bursty manner from the edge and into
the \sol\ region, will eventually be lost by transport along open
field lines. The characteristics time between the bursts is
related to the viscous diffusion (compare with Fig.\ 3 in
Ref.~\cite{garciaetal04}, where the value of $\nu $ is $ 5 \times
10^{-3}$). We further verified that the self-sustained poloidal
flow profiles are strongly sheared in the edge region, and have
larger amplitudes during the strong fluctuation period.

\begin{figure}
%\begin{figure}[H]
\centering
\includegraphics[width=.75\textwidth]{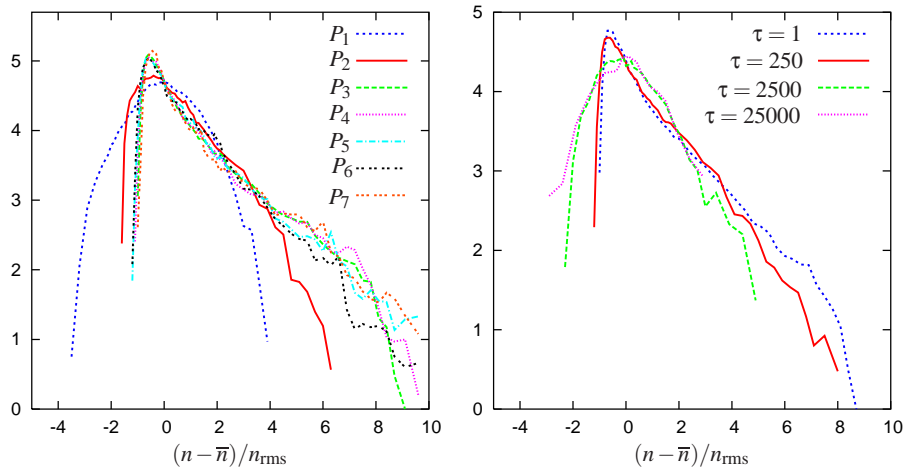}
\caption{In the left panel is shown the probability distribution
functions of particle density measured at seven different radial
positions $P_i$ as shown in Fig.~\ref{fig:geometry}. To the right
is the coarse grained probability distribution function taken at
probe 3, averaged over the time intervals $\tau $ indicated by the
labels. With increasing $\tau $ the skewness decreases as: 2.6,
1.9, 0.73, 0.16, and the flatness factor decreases as: 12.0, 7.8,
3.9, 3.2. For both plots the vertical axis shows count numbers on
a logaritmic scale. $\bar{n}$ designates the averaged density.}
\label{fig:pdf}
\end{figure}

The statistics of single-point recordings at different radial
positions $P_i$ indicated in Fig.~\ref{fig:geometry} agree very
well with experimental measurements. In Fig.~\ref{fig:pdf} we
present the probability distribution functions (histogram of
counts) (\pdf) of the density signals taken from a long-run
simulation containing more than a hundred strong burst events. It
is notably that the PDF at the first probe inside the \lcfs\ is
close to a Gaussian with skewness $ 0.12$ and flatness factor $
2.97$, while the PDF becomes more flat and skewed further into the
\sol. This indicates the high probability of large positive
fluctuations corresponding to blobs of excess plasma. The skewness
and flatness factors grow  through out the \sol\ and take values
up to 4 and 25, respectively. The \pdf's in the \sol\ have similar
structure with a pronounced exponential tail towards large values,
a characteristic feature of turbulent convection in the presence
of sheared flows~\cite{ct:garcia,ct:gb}.

We have also considered the coarse-grained \pdf\, which is
obtained by averaging the signal over time intervals of lengths
$\tau $ and constructing new time records with a time resolution
of $\tau $: $n_{\tau }(t) = (1/\tau )\int_{t - \tau/2}^{t +
\tau/2} n(t^{\prime}) \, dt^{\prime } $. The coarse grained \pdf's
(\pdf$_{\tau } $) for the signal at $P_3 $ are also plotted in
Fig.~\ref{fig:pdf} for increasing values of $\tau $. We observe
that \pdf$_{\tau } $ approaches a Gaussian distribution when $\tau
$ is exceeding the averaged time interval between bursts, which is
roughly $10^4 $. This shows the absence of self-similarity, which
is characteristic for an intermittent signal (see, e.g.,
\cite{carboneetal00}).

\begin{figure}
%\begin{figure}[H]
\centering
\includegraphics[width=.75\textwidth]{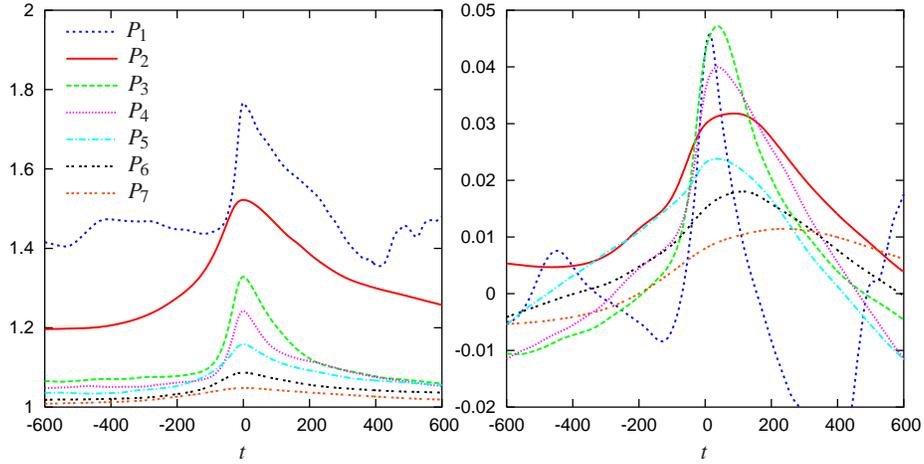}
\caption{Conditionally averaged wave forms of the particle density
(left panel) and the radial velocity $v_x $ measured at seven
different radial positions $P_i$ as shown in
Fig.~\ref{fig:geometry}, using the condition $n(x_{P_i}) -
\bar{n}(x_{P_i}) > 4n_\text{rms}(x_{P_i} )$.}
\label{fig:conditional}
\end{figure}

The conditionally averaged temporal wave forms of the density
calculated from the same signals and the radial velocity field
$v_x $, using the trigger condition $n - \bar{n} > 4n_\text{rms}$
at each individual point, are presented in
Fig.~\ref{fig:conditional}. For the density signal an asymmetric
wave form with a sharp rise and a relatively slow decay is clearly
seen, as also observed in experimental measurements
~\cite{ct:antar,ct:boedo,ct:terry}. The maximum density excursions
significantly exceed the background level, and decay as the
structures propagate through the \sol. By using a negative
amplitude for the conditional averaging very few realizations
results, confirming the presence of blob-like structures. For the
velocity signal we observe that the radial velocity is positive
(directed radially outwards) in the blob. In the edge region it
takes weak negative values both before and after the blob. Also
this result agree with experimental observations ~\cite{ct:boedo}.
We note that the maximum value of $v_x $  decreases on passing the
\lcfs\ and then increases to a maximum value of 0.046 at $P_3$,
after which it slowly decays. From two-dimensional animations we
clearly observe the radial propagation of blob-like structures for
the density and temperature fields, while the vorticity displays a
dipolar structure  as expected from theory and experiment (cf.\
Ref.\ ~\cite{garciaetal04}). From such animations and radial
correlations we find that the radial propagation velocity of the
blob structures corresponds to around $0.05 c_s $ consistent with
Fig.~\ref{fig:conditional}, but with a large statistical variance
in agreement with experimental
measurements~\cite{ct:antar,ct:boedo,ct:terry}.

By combining the conditional evolution of $n $ and $v_x $ in
Fig.~\ref{fig:conditional} we deduce that the blobs are carrying a
large particle flux. We have examined the \pdf\ of the particle
flux averaged over the periodic $y$-direction (the flux surface)
at different radial positions. The \pdf's are quite similar and
strongly skewed with a flat exponential tail towards positive flux
events, showing that the flux is dominated by strong bursts. The
tail of the \pdf\ was found to be well fitted by an extreme value
distribution ~\cite{naulinetal04}. By coarse graining the \pdf\ as
described above we observe a similar behavior as for the local
density fluctuations: the distribution approaches a Gaussian for
large time scales.

We have demonstrated that a two-dimensional model for interchange
turbulence provide results in good agreement with that reported
from experimental investigations of \sol\ turbulence and transient
transport events~\cite{ct:antar,ct:boedo,ct:terry}. An important
feature of the model is the spatial separation between forcing and
damping regions. Our results are in quantitative agreement with
experimental measurements of field-aligned blob-like structures
propagating far into the scrape-off layer. The associated
intermittent transport may have severe consequences for magnetic
confinement experiments by producing large heat bursts on plasma
facing components.\\
\\

This work was supported by the Danish Center for Scientific
Computing through grants no.\ CPU-1101-08 and CPU-1002-17.
O.~E.~Garcia has been supported by financial subvention from the
Research Council of Norway.

%\newpage

\end{document}